\documentclass[fleqn,twoside]{article}

\usepackage{espcrc2}
\usepackage{epsfig}

\title{
QCDml: First milestone for building an International Lattice Data Grid
}
\author{%
  C.M.~Maynard\address[Ed]{%
	School of Physics,
	University of Edinburgh,
	Edinburgh EH9 3JZ, UK},
  D.~Pleiter\address[NIC]{%
	John von Neumann Institute NIC / DESY Zeuthen,
	D-15738 Zeuthen, Germany},
}

\begin{document}

\begin{abstract}
We present an XML schema for marking up gauge configurations called
QCDml. We discuss the general principles and include a tutorial for
how to use the schema.
\end{abstract}

\maketitle

\section{REPORT FROM THE METADATA WORKING GROUP}

To achieve ILDG's aim of sharing gauge field configurations world-wide
a standardised description of configurations is mandatory.  
XML (EXtensible Markup Language) is the language of choice for metadata
since it is designed to describe data.
These metadata documents will be both human readable,
since XML is verbose, and easy to parse by computers. Finally,
standards on the structure and contents of XML documents can be
enforced by using XML schemata.\footnote{See
section~\ref{sec:tutorial} for references and further details.}

The ILDG metadata working group~\cite{ildg-mdwg} addressed in recent
years the task of defining an XML schema. During the 2003 lattice
conference~\cite{ildg-lat03} the group presented an initial proposal.
Since then the strategy for marking-up the physics parameters has been
revised. However, whilst the contents remained unchanged, the
usability has been significantly improved. The working group presented
at this conference the first working version of the schema, QCDml.

Many lattice practitioners, who are typically not familiar using XML
yet, might ask whether the proposed strategy is too complicated.
However, using XML is much easier than many might expect. A large
number of software tools exists for creating and parsing XML
documents.  When looking at the proposed schema, it should be realised
that it's complexity originates from the large variety of different
simulations being carried out within the lattice community. Metadata
documents will only contain information on one particular
simulation. All metadata documents will have to conform to the
schema. It is the schema which contains the complexity which allows
the many different actions being used for simulating QCD with
dynamical fermions.

During the design process three general requirements have been taken
into account. Firstly, the schema has to be \emph{extensible} as
parameters of future simulations cannot be anticipated. This has to be
done in such a way that any metadata document which conforms to the
current schema will also conform to any future extended schema. The
long-term validity of all metadata documents published by users of
ILDG is a definite design goal of the schema. Secondly, the mark-up of
simulation parameters has to be \emph{unique} to avoid, e.g., the same
action being described in two different ways. This would otherwise
spoil the possibility to search for certain configurations.  Finally,
the schema has been kept \emph{general} enough to allow the
description of data other than gauge configurations (propagators,
correlators, etc.) in the future.

\subsection{Overview on the xml schemata}

\begin{figure}[ht]
\vspace*{-5mm}
\begin{center}
\epsfig{file=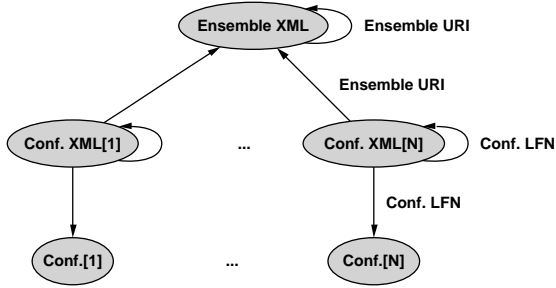,width=75mm}
\end{center}
\vspace*{-10mm}
\caption{Logical file names (LFN) and URIs linking ensemble and 
         configuration XML documents as well as gauge configurations.}
\label{fig:lfn}
\vspace*{-5mm}
\end{figure}

Gauge configurations are generated by a Markov chain. All
configurations from one chain share many properties. Therefore the
metadata can be split into two documents. The \emph{ensemble XML}
document contains all parameters which remain unchanged for the whole
Markov chain.  Other parameters are specific to one or a set of
consecutive Markov steps and will be stored in a \emph{configuration
XML} document. A Universal Resource Indicator (URI) is used to link
these two documents as well as the Logical File Name (LFN) to link
the configuration XML document and the
gauge configuration itself (see Fig.~\ref{fig:lfn}).  For both types
of XML documents corresponding schemata have been developed which can
be downloaded from the working group's web-site~\cite{ildg-mdwg}.

\begin{figure}[ht]
\vspace{-5mm}
\begin{center}
\epsfig{file=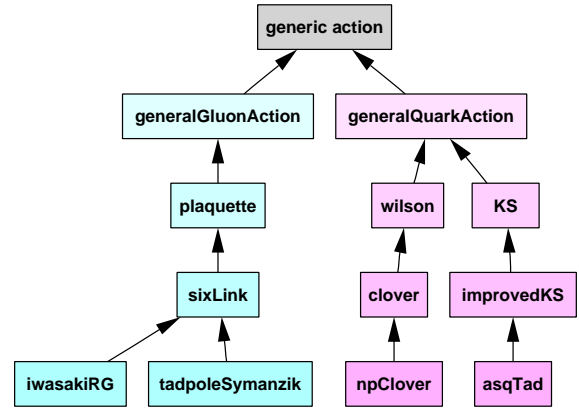,width=75mm}
\end{center}
\vspace*{-10mm}
\caption{Hierarchy of actions in the ensemble XML schema.}
\label{fig:hierarchy}
\vspace*{-5mm}
\end{figure}

An example for parameters which will be the same for all
configurations of an ensemble are the physics parameters. The corresponding
parts of the ensemble XML document consists of information about the lattice
size and a mark-up of the action. The description of the action is most
critical for preserving uniqueness and extensibility of the schema.
The metadata working group adopted the following strategy which
is visualised in Fig.~\ref{fig:hierarchy}:
\begin{itemize}
\item Each action can be split into a gauge and a fermion action.
\item The ensemble XML schema contains an element
      \texttt{<generalGluonAction>} and an optional element
      \texttt{<generalQuarkAction>} which will substituted by
      the actually used action.
\item Actions which contain a structure which is the same as
      for a simpler action are ordered by an inheritance tree.
      For example, the clover fermion action is equivalent to
      the standard Wilson fermion action plus an improvement term.
\item Actions which have the same structure in common are grouped.
      For instance, the Iwasaki and the Symanzik improved gauge
      actions only differ by the choice of the couplings.
\end{itemize}
This inheritance tree of possible actions is obviously extensible. Any
action will be included into the schema only once to ensure uniqueness.

The description of each action is organised in three parts
(See Fig.~\ref{fig:action}). Firstly, an array of \texttt{<couplings>}
allows to store the names and values of all couplings and, in case of
the fermion action, the number of flavours. Secondly, a description of
the fields is required to store information, e.g., about the used
normalisation or boundary conditions. Finally, any further information
can be stored in a \emph{glossary}.

The element \texttt{<glossary>} contains a URL to a document provided
by the contributors. This document does not have to conform to any
schema, it may even be not an XML but rather a human readable
document, e.g.~a TeX file.  This gives the contributors the freedom to
store all kind of information with regard to the used action, for
instance information on the particular choice of
couplings. Nevertheless, some guidelines will be needed to ensure that
these documents contain all relevant information in a comprehensive
form.

The variety of algorithms being used in lattice simulations is even
larger than the number of different actions. The parameters of the
algorithms are therefore essentially unconstrained. It should be
noted that as a consequence such parameters are in practise not
searchable. The only constrained element \texttt{<exact>} provides
information on whether the algorithm being used is exact or not.

It will be mandatory to provide a reference to a publication on the
used algorithm and an URL to a glossary document. Furthermore, all
submitters are strongly encouraged to provide a full list of all
algorithmic parameters used in their simulations.  The names of the
parameters should be chosen in such a way that they can be uniquely
related with the algorithmic parameters described in the publication
and the glossary file. Unlike the physics parameters the algorithmic
parameters might change when generating a Markov chain. For instance,
the step size of the HMC algorithm might be adjusted during a
run. While the ensemble XML document will contain most of the
information on the used algorithm, the submitter can store those
parameters which might change within an ensemble into the
configuration XML document.

As an matter of good scientific research practise, the generation of
each configuration should be fully and comprehensively documented.
Therefore submitters will have to provide information which machine
and what code has been used to generate a particular configuration.
Each machine can be identified by machine (or partition) name, the
hosting institution and the machine type. Additional information can
be stored as an optional comment. Concerning the simulation program
submitters have to ensure that it can be identified by a name,
a version string (e.g.~a CVS tag), and the date of compilation.
Again an optional comment allows to add further information, e.g.~on
compile time variables. All these parameters are not constrained and
therefore not searchable. Only the information on the precision
used to generate configurations will be searchable, as users
might care about the used machine precision, in
particular when quark masses become light.

The metadata will also include information about who submitted a
configuration to ILDG within which project. This information can be
stored in the management section which is foreseen in both the
ensemble and the configuration XML document. Within this section
also information will be stored which allows the user of a configuration
to check the integrity of the downloaded data. To do so he can
verify the checksum for the binary files, which will however not be
preserved when transforming the gauge configuration into a different
format. The user can still perform another test by recalculating
the plaquette value and comparing this with the value stored in the
configuration XML document. It should however be noticed that this
test is less strong as both values will only agree within rounding
errors and because the plaquette value is preserved  by various
transformations of a gauge field configuration.

All operations affecting an ensemble or just a particular configuration
should be documented. Possible actions include the insertion and
modification of an ensemble and the insertion, replacement
or even the revocation of a configuration. The last two actions might
be necessary if for example the computer or the code which was used
to generate a configuration turned out to be broken. It should be
noted that the submitters of configurations might not have to generate
this information themselves, as the user interfaces to be developed
for performing such actions could take care of patching the
ensemble and configuration XML documents accordingly.

\section{QCDML TUTORIAL}\label{sec:tutorial}

The purpose of this section is to demonstrate how to mark up configurations 
according to the XML schema QCDml. We start with some Frequently Asked 
Questions (FAQ) about XML schema.

\subsection{XML Schema FAQ}
\begin{itemize}
  \item What is XML Schema?
  \begin{itemize} \item XML schema is a collection of rules 
                              for XML documents
                  \item An XML schema is itself an XML instance Document
			      (ID)
        \end{itemize}
  \item Why do we need an XML schema?
  \begin{itemize} \item So that computers can read and understand XML IDs
                  \item e.g. \verb.<length>16</length>.
                  \item The meaning of length is context dependent, 
                              the schema makes this information explicit
   \end{itemize} 
   \item Do users need to learn XML schema?
        \begin{itemize} 
             \item No. XML schema makes it easier to write XML IDs
        \end{itemize}
\end{itemize}

\subsection{Getting started}
QCDml1.1 is available for use and can be downloaded along with
documentation and example XML IDs from the ILDG website~\cite{ILDG} by
following the links in the metadata section. In QCDml1.1 the
metadata is split into two parts. Metadata which is common to all
configurations in an ensemble lives in the namespace of the ensemble,
and only one XML ID for the whole ensemble is required. 

An XML namespace is defined by W3C~\cite{w3C:XML} consortium as {\em a
collection of names identified with a URI reference}. 
Metadata which
is specific to each configuration lives in a separate namespace and an
XML ID is required for each configuration. Below is an XML chunk, it
is the start of an example QCDml ID.

\begin{verbatim}
<?xml version="1.0" encoding="UTF-8"?>
<markovChain xmlns="http://www.lqcd.org/
#ildg/QCDml/ensemble1.1" 
xmlns:xsi="http://www.w3.org/2001/
#XMLSchema-instance" 
xsi:schemaLocation="http://www.lqcd.org/
#ildg/QCDml/ensemble1.1 
#www.ph.ed.ac.uk/ukqcd/community/
#the_grid/QCDml1.1/
#QCDml1.1Ensemble.xsd">
 <markovChainURI>
  www.lqcd.org/ildg/ukqcd/ukqcd1
 </markovChainURI>
 +<management/>
 +<physics/>
 +<algorithm/>
</markovChain>
\end{verbatim}

The ``+'' symbol is used to show that there is substructure below the
element, and the \verb.#. symbol is used to indicate line
continuation. The element \verb.<markovChain/>. is the
root of the XML ID. The rest of the first line is the URI which
identifies the namespace of the ensemble metadata. This has no prefix
to identify elements which belong to this namespace as it is the
default namespace. The second line is the namespace of XML schema
itself.  The third and fourth lines give the location of the file
which contains the schema. The attribute
\verb#xsi:schemaLocation# is used to link the URI which identifies the
namespaces with a URL which is the file which contains the schema.
This could be a URL which is the URI of the namespace but it doesn't
have to be.

\begin{figure*}
\begin{center}
\epsfig{file=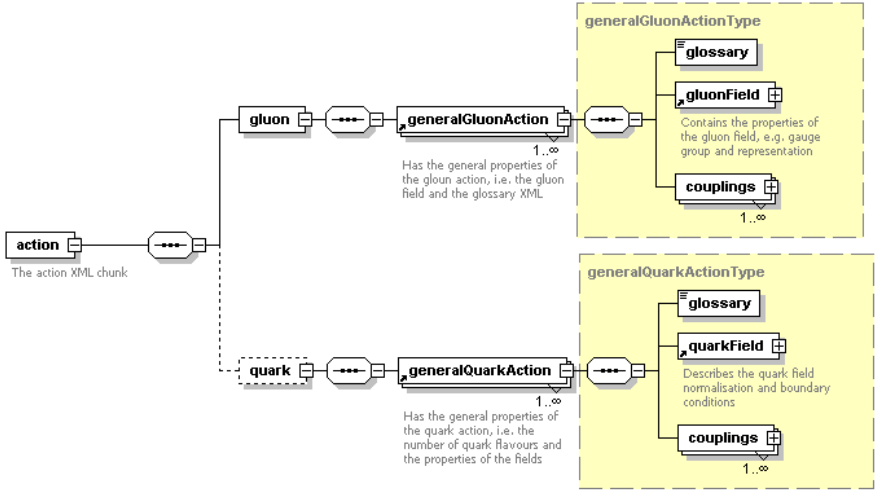,width=125mm}
\caption{A diagram of the action in QCDML}
\label{fig:action}
\end{center}
\end{figure*}

The element \verb#<markovChainURI/># which follows
\verb#<markovChain/># is the URI which identifies this ensemble. Each
configuration XML ID which belongs to this ensemble is linked to it
using this URI.

If an XML ID conforms to the rules of a particular schema it is said
to be {\em valid}. A software application which verifies that an XML
ID is valid is unsurprisingly called a validator. Schema aware
applications can then read and use valid XML IDs. One can write XML
IDs in an editor such as \verb#vi# or \verb#emacs#, however, other
tools are available. XMLspy is commercial software which can be used
for schema and XML ID manipulation, it can, for instance, generate an
XML ID from the schema. There are many other XML manipulation tools,
links can be found at~\cite{W3C:Schema}.

\subsection{Physics and Actions}
The element \verb#<physics/># contains two elements, \verb#<size/># 
and \verb#<action/>#. The former is rather self explanatory and 
contains the size of the system.

Most searches of metadata will be on the action, consequently a lot of
thought has gone into marking up the actions. Some of the object
oriented features of XML schema have been employed in the schema to
categorise actions, such as inheritance and the substitution
group. This enables the XML IDs to be relatively simple. The general
structure is shown in figure \ref{fig:action}. The action has been
split into two parts, gluon and quark. These {\em general} elements
encapsulate the general properties of the actions, such as the fields
and the glossary document. The glossary contains information such as
the mathematical definitions of the actions and a reference to a paper
where the action is discussed. However, this type of information is
not suitable to being marked up in XML, it is essentially
unconstrained and as such is not really searchable by a computer.

\begin{figure}
\begin{center}
\epsfig{file=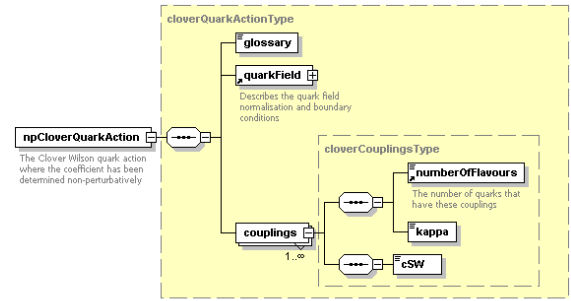,width=75mm}
\vspace*{-10mm}
\caption{A diagram of the NP Clover action}
\label{fig:npClover}
\end{center}
\vspace*{-10mm}
\end{figure}

Specific quark and gluons inherit their properties from the general
actions.  These actions, such as \verb#<wilsonQuarkAction/># have
specific couplings, in this case \verb#<kappa/>#. The
\verb#<cloverQuarkAction/># is an extension of this action, as it is a
Wilson action, but has an extra coupling, \verb#<cSW/>#.  This is
shown in figure \ref{fig:npClover}. An inheritance tree for various
actions can be built up in this way.

\begin{figure}
\begin{center}
\epsfig{file=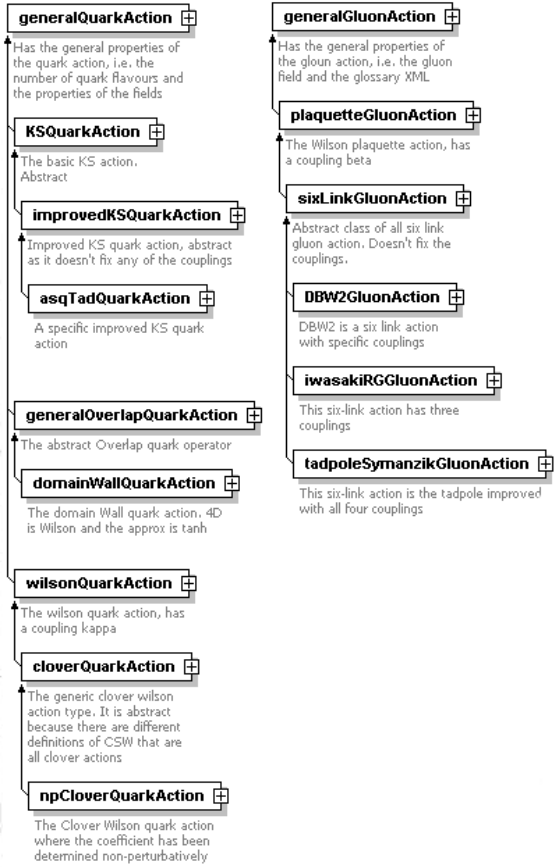,width=75mm}
\vspace*{-10mm}
\caption{A diagram showing the inheritance tree for quark and 
gluon actions.}
\label{fig:actionsInheritance}
\end{center}
\vspace*{-10mm}
\end{figure}

The metadata working group (MDWG) has not set up inheritance trees for
all possible actions, but the schema is extensible so that further
actions can be added without existing XML IDs having to be
modified. Actions that have been added to QCDml are shown at the ILDG
metadata web pages, and an example of which is shown in figure
\ref{fig:actionsInheritance}.

For the gauge actions the metadata working group adopted a particular
convention for \verb#<sixLinkGluonActions/>#
\begin{equation}
  S_g^{\rm 6 link} = \beta\times\left( c_0 {\mathcal P} + c_1 {\mathcal R} 
	+ c_2 {\mathcal C} + c_3{\mathcal X}\right)
\end{equation}
Where $\mathcal{P}$ is the Plaquette Wilson loop, $\mathcal{R}$ the
six-link rectangle, $\mathcal{C}$ the six-link chair and $\mathcal X$
the three dimensional Wilson loop.
The values of some of the couplings can be restricted to certain
ranges or specific values. For example, in the Iwasaki RG action, the
couplings are constrained, $c_2=c_3=0$, $c_0=(1-8c_1)$ and $c_1=-0.331$.

The quark action coupling has an integer valued element
\verb#<numberOfFlavours/>#.  This labels how many flavours have these
couplings, {\em i.e.} how many degenerate flavours. The element
\verb#<couplings/># is array valued, that is this part of the action
can be repeated but with different couplings. This is useful for
marking up non-degenerate quark flavours.

An XML chunk for the $n_f=2$ non-perturbative clover action is shown
below.
\begin{verbatim}
<npCloverQuarkAction>
  <glossary>
    www.lqcd.org/ildg/
#npCloverQuarkAction.xml
  </glossary>
  +<quarkField/>
  <couplings>
    <numberOfFlavours>
      2
    </numberOfFlavours>
    <kappa>0.1350</kappa>
    <cSW>2.0171</cSW>
  </couplings>
</npCloverQuarkAction>
\end{verbatim}
This is quite a
short XML chunk, as the hierarchy npCloverQuarkAction $\rightarrow$
CloverQuarkAction $\rightarrow$ WilsonQuarkAction $\rightarrow$
GeneralQuarkAction is contained in the schema. 

A rather technical point is that in the XPath 1.0~\cite{W3C:XPath}
specification, there is no support for substitution groups which means
that a search for WilsonQuarkAction elements would not return any
cloverQuarkAction elements, although this can be achieved with a
boolean ``or'' such as \texttt{ [/action/quark/npCloverQuarkAction |
/action/quark/WilsonQuarkAction]}. However, the specification for
XPath 2.0 is nearing completion~\cite{W3C:XPath2}, and this issue is
beginning to be addressed.

An XML chunk for the $n_f=2+1$ AsqTad Kogut-Susskind quark action
is shown below.
\begin{verbatim}
<asqTadQuarkAction>
  <glossary>
    www.lqcd.org/lqcd/
#asqTadQuarkAction.xml
  </glossary>
  +<quarkField/>
  <couplings>
    <numberOfFlavours>
      2
    </numberOfFlavours>
    <mass>0.02</mass>
    <cNaik>-0.05713116</cNaik>
    <c1Link>0.625</c1Link>
    <c3Link>-0.08569673</c3Link>
    <c5LinkChair>
      0.02937572
    </c5LinkChair>
    <c7LinkTwist>
      -0.006713076
    </c7LinkTwist>
    <cLepage>-0.1175029</cLepage>
  </couplings>
  <couplings>
    <numberOfFlavours>
       1
    </numberOfFlavours>
    <mass>0.05</mass>
    <cNaik>-0.05713116</cNaik>
    <c1Link>0.625</c1Link>
    <c3Link>-0.08569673</c3Link>
    <c5LinkChair>
      0.02937572
    </c5LinkChair>
    <c7LinkTwist>
      -0.006713076
    </c7LinkTwist>
    <cLepage>-0.1175029</cLepage>
  </couplings>
</asqTadQuarkAction>
\end{verbatim}
The structure is the same, and all the couplings are clearly
shown. The non-degenerate quark masses result in a second
\verb#<couplings/># element, but with different number of flavours and
different mass. It is easy to distinguish between $n_f=2+1$ and
$n_f=3$.

\subsection{Management}
This metadata gives the status of the data that is registered with the
ILDG. In that sense it is created when the data is made public. In
principal this would be generated or ``stamped'' by some ILDG
middleware. As this application does not yet exist, it will have to be
generated ``by hand''.  Below is an example of the management chunk of
XML.

\begin{verbatim}
<management>
  <revisions>1</revisions>
  <collaboration>UKQCD</collaboration>
  <projectName>Clover NF=2</projectName>
  <archiveHistory>
    <elem>
      <revision>1</revision>
        <revisionAction>
          add
        </revisionAction>
        <numberConfigs>
          829
        </numberConfigs>
        <participant>
          <name>Chris Maynard</name>
          <institution>
            University of Edinburgh
          </institution>
        </participant>
        <date>
          2004-04-04T16:20:10Z
        </date>
        <comment>
          This is the time of addition 
        </comment>
    </elem>
  </archiveHistory>
</management>
\end{verbatim}

The \verb#<archiveHistory/># element can have several revisions.
\verb#<revision/># is array valued. An ensemble could have configurations added 
to it, replaced or even removed, if a mistake has been found. So the allowed
values of \verb#<revisionAction># are an enumeration of 
\verb#{add,remove,replace}#.  To discover 
how many configurations are in an ensemble, it is relatively easy to
construct an XPath query to find the number of revisions and then the
number of configurations for each revision.

\subsection{Algorithm}
Algorithmic metadata is split between the ensemble and configuration documents,
as it is possible, for instance, to have different stopping requirements for the 
inverter across the ensemble. The algorithmic metadata is in the form of
unconstrained \verb#<name/> <value/># pairs. For example
\begin{verbatim}
<algorithm>
  <name>GHMC</name>
  <glossary>
    www.ph.ed.ac.uk/ukqcd/
 #community/GHMC.xml
  </glossary>
  <reference>
    Phys.Rev.D65:054502,2002
  </reference>
  <exact>true</exact>
  <parameters>
    <name>stepSize</name>
    <value>0.00625</value>
  </parameters>
</algorithm>
\end{verbatim}

It would be very difficult to create a hierarchical structure for
algorithms, and especially difficult to make such hierarchy
extensible. Again there is a glossary document which contains the free
text, or mathematical definition of the algorithm, and a reference to
a paper which describes the algorithm. There is also the boolean
valued element \verb#<exact/># which denotes whether or not the
algorithm is exact.

\subsection{Configuration XML}
The configuration XML follows along similar lines. However, it is much
shorter and so in principle could be directly output from the code
that produced the configuration. Below is an example configuration XML
ID.  Again we start with a set of namespace declarations, which whilst
the default namespace for configuration is separate from that of the
ensemble, it still follows the same pattern.

The management section is very similar to that of the ensemble,
however, there is an important addition: there is a ``zeroth''
revision which is {\em generate}. There is important metadata of when
the gauge configuration was generated, and not just when it is
submitted to the ILDG catalogue. As noted above ILDG middleware will
eventually create the management part of the metadata when it is added
to the ILDG catalogue, but this has yet to be written.  The second
important difference between the ensemble and configuration metadata
is the \verb#<crcCheckSum/># which can be used to verify the data has
been copied correctly.

\begin{verbatim}
<?xml version="1.0" encoding="UTF-8"?>
<gaugeConfiguration 
xmlns="http://www.lqcd.org/ildg/QCDml/
#config1.1"
xmlns:xsi="http://www.w3.org/2001/
#XMLSchema-instance"
xsi:schemaLocation="http://
#www.lqcd.org/ildg/QCDml/config1.1
www.ph.ed.ac.uk/ukqcd/community/
#the_grid/QCDml1.1/QCDml1.1Config.xsd">
  <management>
     <revisions>1</revisions>
    <crcCheckSum>           
      2632843688
    </crcCheckSum>
    <archiveHistory>
      <elem>
        <revision>0</revision>
        <revisionAction>
          generate
        </revisionAction>
        <participant>
          <name>Chris Maynard</name>
          <institution>
            Edinburgh
          </institution>
        </participant>
        <date>
          1998-04-24T10:25:52Z
        </date>
      </elem>
      <elem>
        <revision>1</revision>
        <revisionAction>
          add
        </revisionAction>
        <participant>
          <name>Chris Maynard</name>
          <institution>
            University of Edinburgh
          </institution>
        </participant>
        <date>
          2002-04-24T10:25:52Z
        </date>
      </elem>
    </archiveHistory>
  </management>
  <implementation>
    <machine>
      <name>T3E-900</name>
      <institution>
        epcc Edinburgh
      </institution>
      <machineType>
        Alpha processor
      </machineType>
    </machine>
    <code>
      <name>
        UKQCD FORTRAN
      </name>
      <version>16.8.3.1</version>
      <date>
        1997-04-04T16:20:10Z
      </date>
    </code>
  </implementation>
  <algorithm>
    <parameters>
      <name>targetResidue</name>
      <value>1e-07</value>
    </parameters>
  </algorithm>
  <precision>single</precision>
  <markovStep>
    <markovChainURI>www.lqcd.org/
#ildg/ukqcd/ukqcd1</markovChainURI>
    <series>1</series>
    <update>010170</update>
    <avePlaquette>
      0.53380336E+00
    </avePlaquette>
    <dataLFN>
      D52C202K3500U010170
    </dataLFN>
  </markovStep>
</gaugeConfiguration>
\end{verbatim}

The next element is \verb#<implementation/># which holds information
such as code versions, and machine version. Both of these entries are
really only important for bug tracking, but if ever a bug is found
then they are vital for tracking down the effected
configurations. This metadata section is best written by the code that
generated the configuration, as it is quite easy for this metadata to
become lost.

The \verb#<algorithm/># element is the same as that of the ensemble,
e.g. a name value pair for each algorithmic parameter that is specific
to that configuration. The \verb#<precision/># element is also
algorithmic in nature.  it is the precision in which the configuration
was computed, not in which the data is stored. It is an enumeration of
\verb#{single,double,mixed}#, it is possible to have some parts of
gauge configuration generation code in single precision and some in
double.

The final segment \verb#markovStep# is the most immediately useful. 
\verb#<markovChainURI/># is the URI of the Markov Chain to which this
configuration belongs. This links the ensemble and the configuration
XML IDs together. \verb#<series/># and \verb#<update/># locate the
configuration in the Markov Chain. The average Plaquette is useful for
checking that downloads, copies or data reads have all worked
correctly, not least as this metadata is data format
independent. Finally \verb#<dataLFN/># is the logical filename of the
data on the grid. This links the metadata to the data.  In QCDgrid
(UKQCD's data grid) the data submission tool reads this element from
the metadata and then uses this as the logical file name.
 
This tutorial hopefully gives a flavour of how to mark up gauge
configurations in QCDml1.1. The ILDG website contains more detailed
documentation on the schema along with example XML IDs. The website
will be updated regularly as changes and extensions occur, but this
should still serve as a guide.

\section{FUTURE PROGRESS}
The MDWG along with the middleware working group is actively
considering the issue of data and file formats, but this is discussed
elsewhere. Completing the hierarchy tree for all commonly used
actions is another task to be finished.
Gauge configurations are not the only data that could be
shared by ILDG members, for instance quark propagators and hadron
correlator. The MDWG is considering how to extend QCDml to such data.

\end{document}